\documentclass[12pt,%
 reprint,
 amsmath,amssymb,
 aps,
]{revtex4-1}

\usepackage{graphicx}
\usepackage{dcolumn}
\usepackage{bm}
\usepackage{subfigure}
\usepackage{times}
\usepackage[bookmarks=true,colorlinks=true,linkcolor=blue,
urlcolor=blue,citecolor=blue]{hyperref}

\begin{document}

\preprint{APS/123-QED}

\title{\bf Excited states of the quasi-one-dimensional hexagonal quantum antiferromagnets}

\author{M. Merdan$^{1,2}$ and Y. Xian$^1$}
 \affiliation{%
 {$^1$School of Physics and Astronomy, The University of Manchester, Manchester M13 9PL, UK}\\
$^2$Department of Physics, College of Science, University of Babylon, Babylon, Iraq }
\date{\today}

\begin{abstract}
We investigate the excited states of the quasi-one-dimensional quantum antiferromagnets on hexagonal lattices, including the longitudinal modes based on the magnon-density waves. A model Hamiltonian with a uniaxial single-ion anisotropy is first studied by a spin-wave theory based on the one-boson method; the ground state thus obtained is employed for the study of the longitudinal modes. The full energy spectra of both the transverse modes (i.e., magnons) and the longitudinal modes are obtained as functions of the nearest-neighbor coupling and the anisotropy constants. We have found two longitudinal modes due to the non-collinear nature of the triangular antiferromagnetic order, similar to that of the phenomenological field theory approach by Affleck. The excitation energy gaps due to the anisotropy and the energy gaps of the longitudinal modes without anisotropy are then investigated. We then compare our results for the longitudinal energy gaps at the magnetic wavevectors with the experimental results for several antiferromagnetic compounds with both integer and non-integer spin quantum numbers, and we find good agreement after the higher-order contributions are included in our calculations.

\begin{description}
\item[PACS numbers]
75.10.Jm, 75.30.DS, 75.50.Ee.
\end{description}
\end{abstract}

\pacs{Valid PACS appear here}
\maketitle

\section {Introduction}

The excitations of the quasi-one-dimensional (1d) Heisenberg antiferromagnets systems have been studied extensively since Haldane predicted an energy gap in the excitation spectra of the isotropic \emph{integer}-spin Heisenberg chains in 1983 \cite{PhysRevLett.50.1153}. Now it is well established that there is an energy gap separating the singlet ground state from the triplet lowest-energy-excitation states for the \emph{integer}-spin Heisenberg chains, contrast to the gapless excitation states of the \emph{half-odd-integer}-spin Heisenberg systems \cite{PhysRevB.52.13368,PhysRevB.24.1429}. This theoretical prediction has been confirmed by Buyers \emph{et al} \cite{PhysRevLett.56.371} in the inelastic-neutron-scattering experiments on the quasi-1D antiferromangetic compound CsNiCl$_3$. Some subsequent experimental investigations \cite{PhysRevB.50.9174,PhysRevLett.56.371, PhysRevLett.69.3571,Steiner1987,PhysRevLett.87.017201} and numerical calculations \cite{PhysRevB.49.13235,PhysRevB.46.10854,
PhysRevLett.75.3348,PhysRevB.48.10227,PhysRevLett.62.2313} also support Haldane's prediction.

At very low temperature, most of the quasi-1D antiferromagnetic materials including CsNiCl$_3$ show the three-dimensional nature with the classical magnetic order, and more interestingly, energy gaps at the magnetic wavevector have also been observed in many compounds \cite{PhysRevLett.56.371}. For the case of CsNiCl$_3$, the observed energy gap was initially explained by a uniaxial single-ion anisotropy but now it is widely accepted that the gapped excited state belongs to one of the two longitudinal modes corresponding to the oscillations in the magnitude of the magnetic order of the quasi-1D hexagonal systems, first proposed by Affleck based on a simplified version of Haldane's theory \cite{Affleck1989,PhysRevB.46.8934}. The gapped longitudinal modes are clearly beyond the conventional spin-wave theory which produces only the transverse excitations usually referred to as magnons. A Later experimental study by Enderle \emph{et al.} \cite{PhysRevB.59.4235} using high-resolution polarized neutron scattering also confirms Affleck's proposal of the longitudinal modes, and contradicts the spin-wave theory of two-magnon by Ohyama and Shiba \cite{JPSJ.62.3277} or a modified spin-wave theory by Plumer  and Caill\'e \cite{Plumer1992}. There are also investigations of the longitudinal excitation states in other quasi-1D structures with the N\'eel-like long-ranged order at low temperature such as the tetragonal KCuF${}_3$ with $s=1/2$ \cite{lake2005}, where good agreements between the experiment and a theory based on a sine-Gordon field theory have been found for the energy gap at the magnetic wavevector \cite{Schulz1996,Essler1997}. More recently, a longitudinal mode was also observed in the dimerized antiferromagnetic compound TlCuCl${}_3$ under pressure with a long-ranged N\'eel order \cite{Ruegg2008}.

We recently proposed a general microscopic many-body theory based on the magnon-density waves for the longitudinal excitations of spin-$s$ quantum antiferromagnetic systems \cite{Xian2006,Xian2007}. In analogy to Feynmann's theory of the low-lying excited states in the helium-4 superfluid \cite{Feynman1954,Feynman1956}, we identify the longitudinal excitation states in a quantum antiferromagnet with a N\'eel-like order as the collective modes of the magnon-density waves. In application to the quasi-1D tetragonal structure of KCuF${}_3$ with $s=1/2$, with no other fitting parameters than the nearest-neighbor coupling constants in the model Hamiltonian, we find that our numerical results for the energy gap values at the magnetic wavevector are in general agreement with the experiments \cite{yang2011}. We hope that more experimental results for the energy spectra at other wavevectors will be available for comparison.

In this article, we extend our microscopic approach to the quasi-1D hexagonal quantum antiferromagnets such as CsNiCl${}_3$ and RbNiCl${}_3$ \cite{PhysRevB.10.4643,PhysRevB.59.4235,PhysRevB.70.224420} both with spin-1 and CsMnI${}_3$ with spin-5/2 \cite{PhysRevB.43.679}. Furthermore, we also investigate the higher-order contributions to the longitudinal excitation spectra in the large-$s$ expansion. The basal planes of these materials are antieferromagnetic triangular lattice with the noncollinear magnetic order. Hence there are two possible longitudinal modes in these hexagonal systems, rather than the single longitudinal mode of the bipartite systems such as the tetragonal KCuF${}_3$. Some preliminary results for the two dimensional triangular model have been published \cite{M.Merdan2012}. We organize this article as follows. For completeness, we outline the main results of the spin-wave theory for the quasi-1D model in Sec.~II, using the one-boson approach after two spin rotations. We obtain the full spin-wave spectra as a function of the uniaxial single-ion anisotropy. To our knowledge, this anisotropy dependence of the spin-wave spectra has not been published before. We then apply our microscopic theory for the longitudinal excitations in Sec.~III, using the approximated ground state from the spin-wave theory. The energy gaps due to the anisotropy and the energy gaps of the longitudinal modes without anisotropy are then discussed in details. We compare our results for the longitudinal energy gaps with the experimental results for the spin-1 compounds CsNiCl$_3$ and RbNiCl$_3$ and the spin-5/2 compound CsMnI${_3}$. We find good agreements for the energy gap values for CsNiCl$_3$ and RbNiCl$_3$ after including the higher-order contributions in our calculations. For CsMnI${_3}$ which is very close to the pure 1d system we find a big discrepancy between our approximation of the gap value and the experimental results. We conclude this article by a summary and a discussion of the possible further corrections particularly for CsMnI${_3}$.

\section{The spin-wave theory of the anisotropic hexagonal antiferromagnetic systems}

\begin{figure}[h!]
\centering
\hspace{-0.6cm}
\subfigure[]{
   \includegraphics[scale=1.1]{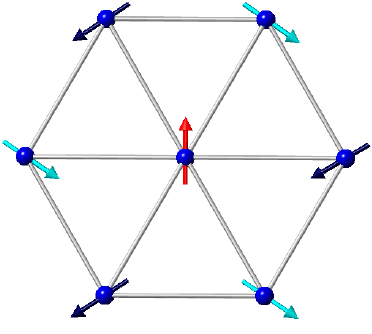}
 }
 \hspace{0.4cm}
 \subfigure[]{
   \includegraphics[scale=1]{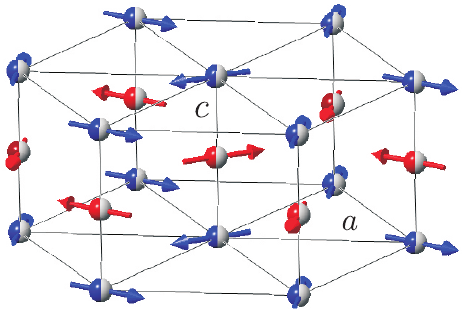}
}
\caption{(Color online)The classical spin structure of the quasi-1D hexagonal antiferromagnets: (a) on the $ab$ plane, and (b) the three-dimensional structure.}
\label{fig1}
\end{figure}

The quasi-1D materials such as CsNiCl$_3$ crystallize in the hexagonal $ABX_3$ structure with space group $P6_3/mmc$, where $A$ is an alkaline-metal cation, $B$ is a cation of the 3$d$ group, and $X$ is a halogen anion. The magnetic ions $B$ constructs the hexagonal lattice in the $ab$ plane with adjacent spins forming angles of $\theta=2\pi/3$, and antiparallel adjacent spins along the chain of the $c$ axis as shown in Figs.~\ref{fig1}(a) and (b). The lattice constants of CsNiCl$_3$, for example, are $a=7.14$ ${\buildrel _{\circ} \over {\mathrm{A}}}$ and $c=5.90$ ${\buildrel _{\circ} \over {\mathrm{A}}}$, and the magnetic moments are carried by Ni$^{2+}$. The superexchange interaction between $B$ (Ni$^{2+}$) ions is modeled by an $N$-spin Heisenberg Hamiltonian with a strong intrachain interaction $J$ and weak interchain interaction $J'$ such as
\begin{equation}\label{eq1}
    H=2J\sum_{\langle i,j\rangle}^\text{chain}\mathbf{S}_i\cdot \mathbf{S}_j+2J'\sum_{\langle i,j\rangle}^\text{plane}\mathbf{S}_i\cdot \mathbf{S}_j+D\sum_i(S_i^z)^2,
\end{equation}
where the notation $\langle i,j\rangle$ indicates the nearest-neighbor couplings only and where we have also added an Ising-like single-ion anisotropy term with constant $D (<0)$. Most of the intrachain couplings in $ABX_3$ compounds are antiferromagnetic such as in CsNiCl$_3$ or RbNiCl$_3$ with easy single-site anisotropy, or CsMnBr$_3$ and RbMnBr$_3$ with hard anisotropy \cite{PhysRevB.56.5373,PhysRevB.54.6327}. These intrachain couplings can also be ferromagnetic (i.e., $J<0$) as in CsNiF$_3$ \cite{PhysRevB.44.11773,PhysRevB.54.12932} or CsCuCl$_3$ \cite{Rastelli1994}. We consider only the antiferromagnetic couplings here. Therefore, the classical ground state of each linear chain along the $c$ axis (also denoted as $y$-axis) is a N\'eel state with alternating spin-up (blue) and spin-down (red) alignments as shown in Fig.~1 (b).

We consider a spin-wave theory for the Hamiltonian~(1) based on the one-boson approach by performing two spin rotations. Firstly, we rotate the local axes of all up-spins (blue) by $180^\circ$ so that all spins along each chain align in the same down direction. This is equivalent to the transformation
\begin{equation}
S^\mp_i\rightarrow-S^\pm_i,\quad
S^z_i\rightarrow-S^z_i
\end{equation}
for the first terms in Eq.~(1), leaving the last two terms unchanged. The second rotation is on the hexagonal lattice of the $ab$ plane (or $xz$-plane) on the second and third terms of Eq.~(1). Following Singh and Huse \cite{Singh1} and Miyake \cite{Miyake1992}, for every triangle of the hexagonal lattices [see Fig~1(a)], we rotate the local axes of two spins along the classical direction in the $xz$-plane to align with that of the third spin \cite{Chubukov1994, Chernyshev2009a}. This is equivalent to the rotation of the $i$-sites of Eq.~(1) by the following transformation
\begin{equation}\label{eq3}
\begin{split}
S_i^x&\rightarrow S_i^x\cos(\theta_i)+S_i^z\sin(\theta_i),\\
S_i^y&\rightarrow S_i^y,\\
S_i^z&\rightarrow S_i^z\cos(\theta_i)-S_i^x\sin(\theta_i),
\end{split}
\end{equation}
where $\theta_i\equiv\mathbf{Q_z}\cdot\mathbf{r}_i$ and $\mathbf{Q_z}=(4\pi/3,0,q_z)$ with $\mathbf Q_z$ at $q_z=\pi$ defined as the magnetic-ordering wavevector of the quasi-1D hexagonal systems. The Hamiltonian~\eqref{eq1} after these two transformations is given as
\onecolumngrid
\begin{equation}\label{eq5}
\begin{split}
H=-\frac{1}{2}J\sum\limits_{l,\varrho}^{\text{chain}}
\big[S_l^+S_{l+\varrho}^+&+S_l^-S_{l+\varrho}^-+2S_l^z
S_{l+\varrho}^z]-\frac{1}{2}J'\sum\limits_{l,\varrho'}^
{\text{plane}}\big[S_l^z
S_{l+\varrho'}^z
+\frac{3}{4}(S_l^+S_{l+\varrho'}^++S_l^-S_{l+\varrho'}^-)\\
&-\frac{1}{4}(S_l^+S_{l+\varrho'}^-+S_l^-S_{l+\varrho'}^+)
-2\sin(\theta_l-\theta_{l+\varrho'})
(S_l^zS_{l+\varrho'}^x-S_l^xS_{l+\varrho'}^z)\big]+\tilde{\cal H}^D,
\end{split}
\end{equation}
\twocolumngrid
\noindent where $l$ runs through all sites, $\varrho$ and $\varrho'$ are the nearest neighbor index vectors with coordination numbers $z=2$ along the chain and $z'=6$ on the hexagonal basal planes respectively, and $\tilde{\cal H}^D$ is the rotated anisotropy term. Care should be taken for the two rotations on this anisotropy term. The first rotation of Eq.~(2) leaves it unchanged due to its quadratic form as mentioned before. In order to perform the second rotation of Eq.~(3) involving rotations of the axes of the two spins to align with the axis of the third spin on the triangular planes, we rewrite the anisotropy term of Eq.~(1) in the following equivalent, suitable form
\begin{equation}\label{eq6}
 \sum_i(S^z_i)^2=\frac{1}{z'}\sum_{l,\varrho'}\left[\frac{1}{3}(S_l^z)^2+\frac
 {2}{3}(S_{l+\varrho'}^z)^2\right].
\end{equation}
The transformation of Eq.~\eqref{eq3} to the second term in Eq.~(5) gives
\begin{equation}\label{eq7}
\begin{split}
 \tilde{\cal H}^D=&\frac{1}{z'}\sum_{l,\varrho'}[\frac{1}{3}
 D(S_l^z)^2+\frac
 {2}{3}D[(S_{l+\varrho'}^z)^2\cos^2\theta_{l+\varrho'}\\
 &+(S_{l+\varrho'}^x)^2\sin^2\theta_{l+\varrho'}
 -\cos\theta_{l+\varrho'}\sin\theta_{l+\varrho'}
 (S_{l+\varrho'}^zS_{l+\varrho'}^x\\
 &\quad\quad\quad\quad
 \quad\quad\quad\quad
 \quad+S_{l+\varrho'}^xS_{l+\varrho'}^z)].
 \end{split}
\end{equation}
We notice that this anisotropy form is different from the simple form of Ref.~\cite{Feile1984435} or that of Ref.~\cite{Welz1993}. We believe that Eq.~(6) is the correct form suitable for the hexagonal systems. The energy gaps in the energy spectra due to this anisotropy term will be presented later.

Using the canonical Holstein-Primakoff transformations, the spin operators are expressed in terms of a single set of boson operators $a^\dagger$ and $a$ as,
\begin{equation}\label{eq8}
S^+=\sqrt{2s}fa,\quad S^-=\sqrt{2s}a^\dagger f,\quad S^z=s-a^\dagger a,
\end{equation}
where $f=\sqrt{1-a^\dagger a/2s}\,$ and $s$ is the spin quantum number. The Hamiltonian of Eq.~\eqref{eq5} can then be written as, after Fourier transformations of the boson operators with the Fourier component operators $a_q$ and $a^\dagger_q$ and to the order of $(2s)$,
\begin{equation}
H\approx H_0+H_2,
\end{equation}
where $H_0$ is the classical energy,
\begin{equation}\label{eq9}
    H_0=-2JNs^2-3J'Ns^2+\frac{1}{3}DNs^2
    (1+2\cos^2\theta +\frac{1}{s}\sin^2\theta)
\end{equation}
with $\theta=2\pi/3$ and $H_2$ is given by the quadratic terms in the boson operators as
\begin{equation}\label{eq10}
H_2=s\sum_q\big[A_qa_q^\dagger a_{-q}-\frac{1}{2}B_q(a_q^\dagger a_{-q}^\dagger+a_{q}a_{-q})\big]
\end{equation}
with constants $A_q$ and $B_q$ defined by
\begin{equation}\label{eq11}
\begin{split}
&A_q=4J+6J'(1+\frac{1}{2}\gamma_q)-\frac{2}{3}D(1+2\cos^2
\theta-\sin^2\theta),\\
&B_q=4J\cos q_z+9J'\gamma_q-\frac{2}{3}D\sin^2\theta,
\end{split}
\end{equation}
and $\gamma_q$ defined as usual by
\begin{equation}\label{eq12}
\gamma_q=\frac{1}{z'}\sum_{\varrho'} e^{i\mathbf {q\cdot r}_{\varrho'}}=\frac{1}{3}\Big(\cos q_x+2\cos
\frac{q_x}{2}\cos\frac{\sqrt{3}}{2}q_y\Big).
\end{equation}
The quadratic Hamiltonian $H_2$ of Eq.~\eqref{eq10} is diagonalized by the usual Bogoliubov transformation and can be written in terms of the new boson operators $\alpha_q$ and $\alpha^\dagger_q$ as,
\begin{equation}\label{eq13}
H_2=\Delta H_0+\sum_q{\cal E}_q\left(\alpha_q^\dagger \alpha_q
 +\frac{1}{2}\right),
\end{equation}
where $\Delta H_0$ is the quantum correction to the classical ground state energy of Eq.~\eqref{eq9},
\begin{equation}
\Delta H_0=-2JNs-3J'Ns+\frac{1}{3}
DNs(1+2\cos^2\theta-\sin^2\theta),
\end{equation}
and ${\cal E}_q$ is the spin-wave excitation spectra,
\begin{equation}\label{eq14}
{\cal E}_q=s\sqrt{A_q^2-B_q^2}.
\end{equation}
\begin{figure}[h!]
\centering
\subfigure[]{
\hspace{0.90cm}
   \includegraphics[scale=0.6]{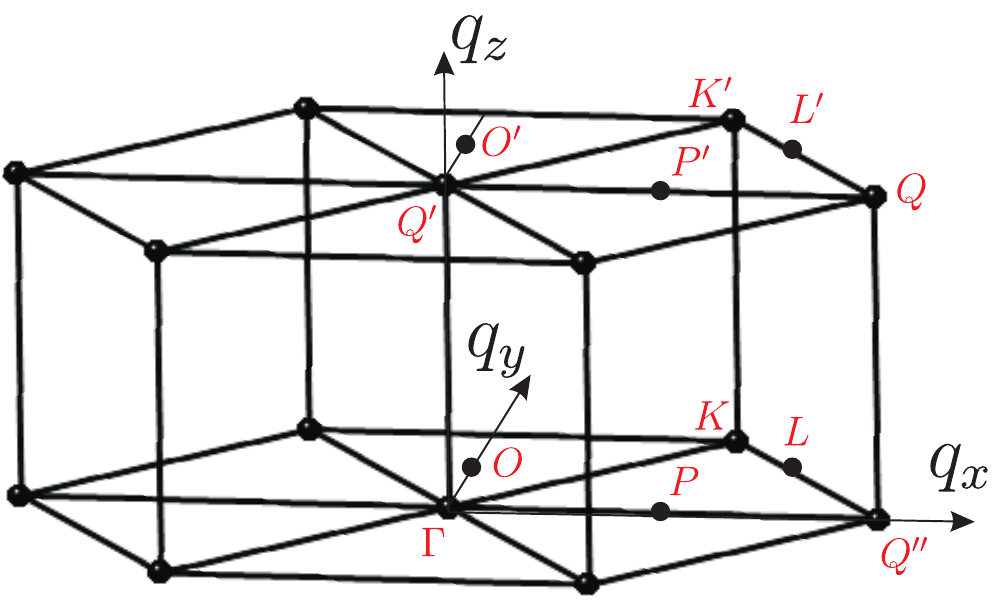}
 }
 \subfigure[]{
   \includegraphics[scale=0.6]{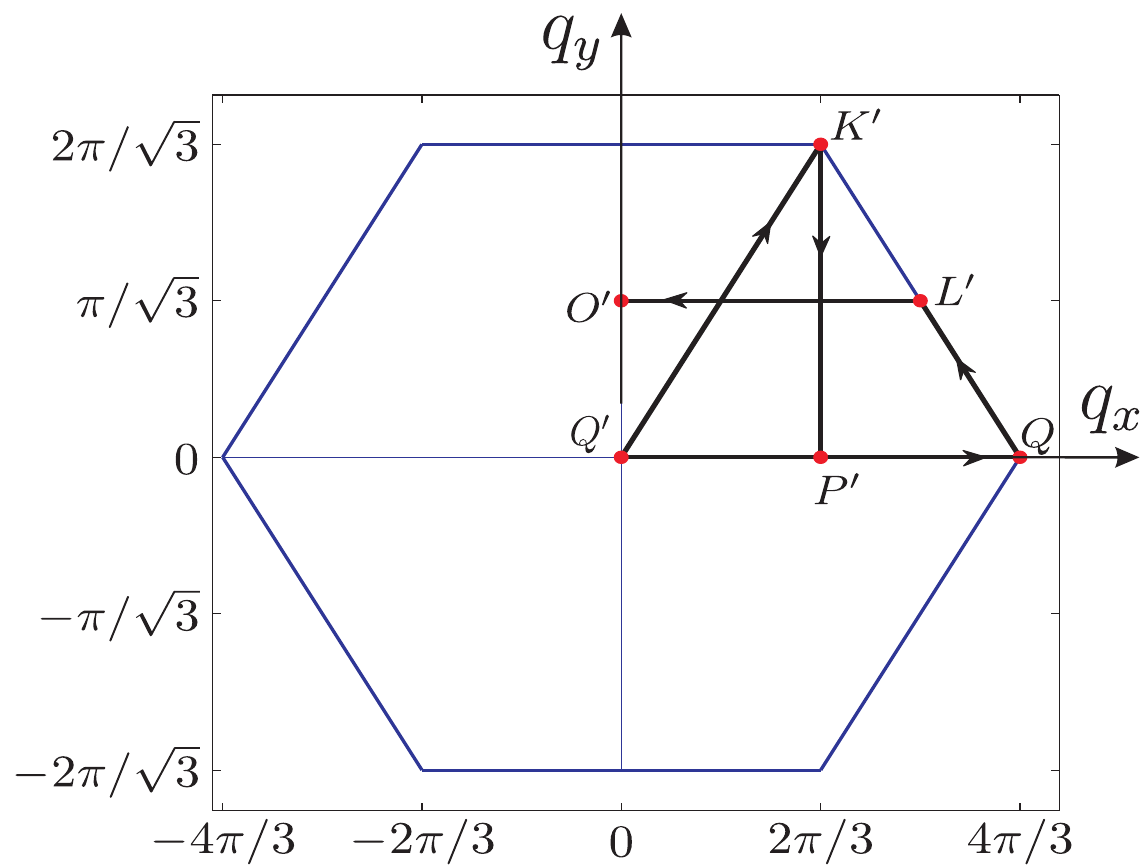}
}
\caption{\small (a) The first Brillouin zone of a quasi-1D hexagonal antiferromagnets. The points  $(0,0)$, $(2\pi/3,2\pi/\sqrt3)$, $(2\pi/3,0)$, $(4\pi/3,0)$, $(\pi,\pi/\sqrt3)$, and $(0,\pi/\sqrt3)$ all at $q_z=\pi$ are denoted as $Q', K', P', Q, L', O'$ respectively, and the similar points at $q_z=0$ are denoted as $\Gamma, K, P, Q'', L, O$ respectively. (b) The hexagonal Brillouin zone at $q_z=\pi$ with some symmetry points in conventional notations for the quasi-1D systems.}
\label{fig2}
\end{figure}

\begin{figure}
\centering
\includegraphics[scale=0.5]{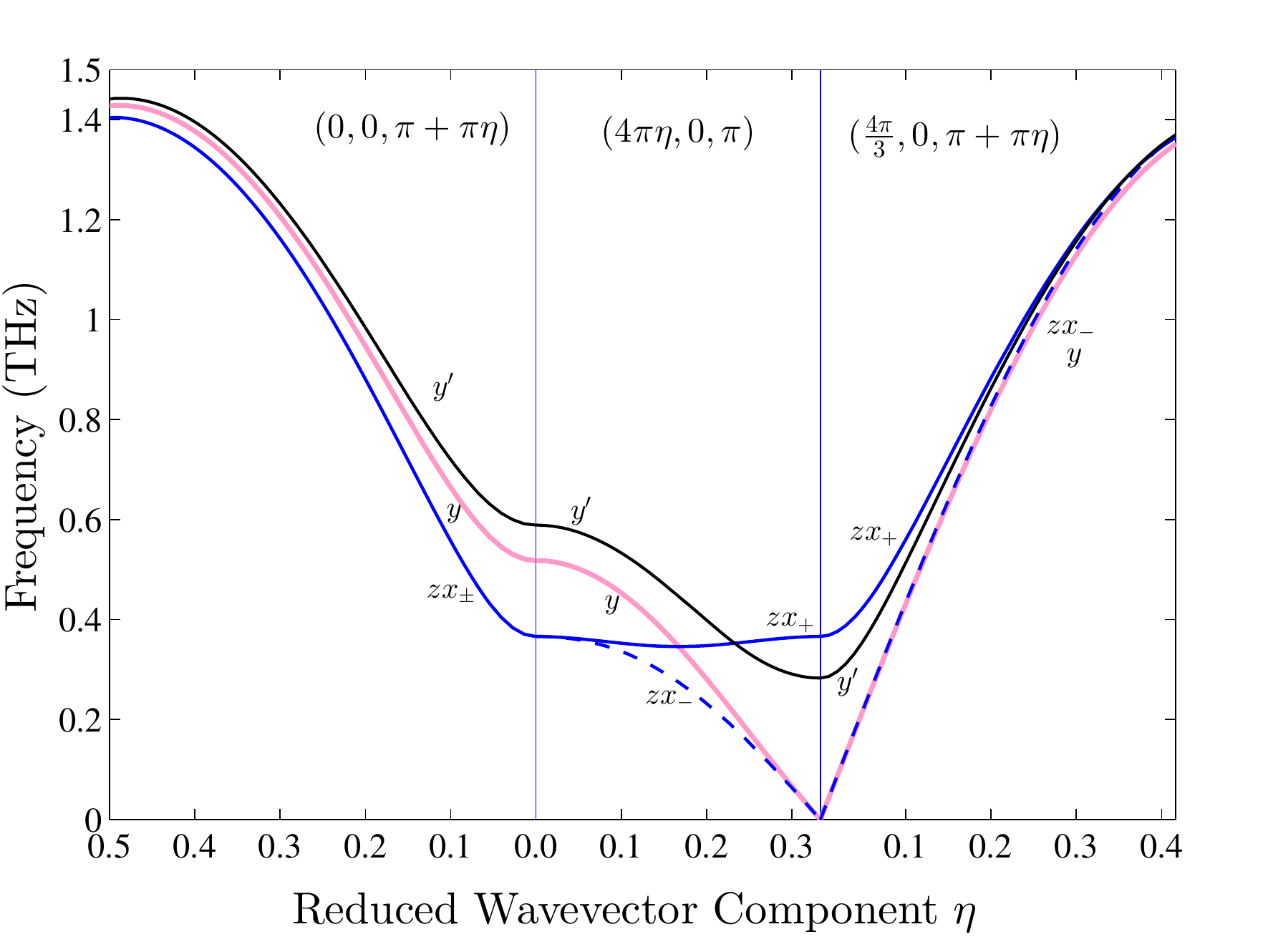}
\caption{The three spin-wave excitation spectra (in colors) for CsNiCl$_3$ with $J=0.345$, $J'=0.0054$ and $D=0$ THz, along the symmetry direction $(0,0,\pi+\pi \eta)$, $(4\pi \eta,0,\pi)$ and $(\frac{4\pi}{3},0,\pi+\pi \eta)$. Also included is the gapped $y$-mode (black, denoted as $y'$) with $D=-0.0285$ using the anisotropy term of Eq.~(6). The solid and dash with the blue color on the lines indicate the $zx_+$-mode and $zx_-$-mode respectively.}
\label{fig3}
\end{figure}

The first Brillouin zone of a quasi-1D antiferromagnet is ploted in Fig.~2, where the magnetic wavevector ${\bf Q}=(4\pi/3,0,\pi)$ is located at the corner of the hexagon and where other symmetry points in conventional notations are also illustrated. We plot the spin-wave spectra of Eq.~(15) in Fig.~\ref{fig3} for CsNiCl$_3$ using the experimental values $J=0.345$, $J'=0.0054$ THz and negligible anisotropy $D\approx0$  \cite{PhysRevLett.56.371,Affleck1989,PhysRevLett.65.2835.2,
0953-8984-3-6-008,PhysRevB.59.4235}. We define the ratio of the two nearest-neighbor coupling constants as $\xi$ and, for CsNiCl$_3$,
\begin{equation}
\xi = \frac{J'}J =0.0157.
\end{equation}
The spin-wave energy spectra with different polarizations are obtained by folding of the wavevectors. In Fig.~3, several branches along the symmetry direction of $(0,0,\eta+1),(\eta,\eta,1)$, and $(1/3,1/3,1+\eta)$ are shown, where $\eta$ is the reduced wave vector component in the reciprocal lattice unit (r.l.u) with $q_z=(2\pi l/c)\cdot(c/2)=\pi l$, and $\gamma=1/3[\cos2\pi h+\cos2\pi k+\cos2\pi(h+k)]$. Using Eq.~(\ref{eq12}) the moving in the paramagnetic Brillouin zone can be written as for $q_x=4\pi\eta$ and $q_z=\pi+\pi\eta$, and the corresponding symmetry directions to those in reciprocal lattice unit are $(0,0,\pi+\pi\eta),(4\pi\eta,0,\pi)$ and $(4\pi/3,0,\pi+\pi\eta)$ respectively. The three transverse spin-wave branches are obtained from Eq.~\eqref{eq14} as follows. The $y$-mode has the polarization along the $y$-axis of the hexagonal lattice where the quantum fluctuation is at $q$; the other two modes are found in the $xz$-plane by translating the wavevector by a magnetic wavevector as $q\to(q\pm Q)$ and are denoted as $zx_\pm$ respectively.

As can be seen from Fig.~3, at the magnetic wavevector $\bf Q$, the $y$-mode is gapless for zero anisotropy ($D=0$). However, as mentioned earlier, an energy gap about $0.41(2J)$ has been observed by the neutron scattering experiments for CsNiCl${}_3$ \cite{PhysRevLett.56.371}. This energy gap can be reproduced in the $y$-mode excitation by introducing an anisotropy with $D=-0.0285$ using our approximation of Eq.~(6), also plotted in Fig.~3. If we use the simple form of Ref.~\cite{Feile1984435} corresponding to setting $\theta=0$ in Eq.(6), the required anisotropy is reduced by a little more than half with the value $D=-0.0141$. Both of these values are now considered too large for CsNiCl${}_3$ which has negligible anisotropy. The conclusion is that the observed gaps are not of the transverse spin-wave spectra, but belong to the longitudinal modes, as first proposed by Affleck \cite{Affleck1989,PhysRevB.46.8934}.

Now we turn our attention to the order parameter. The long-range order of the quasi-1D hexagonal systems is given by the three sublattice-magnetizations with the same magnitude but different orientations as shown in Fig.~1, and it is clearly noncollinear, contrast to the collinear case of the bipartite systems. In the spin-wave theory with one boson method as described above, the magnitude of the sublattice magnetization can be expressed as
\begin{equation}\label{eq17}
    M=\frac{1}{N}\sum_l\langle S_l^z \rangle=s-\rho,
\end{equation}
where the quantum correction $\rho$ is the magnon density defined as the ground-state expectation value of the boson number operator
\begin{equation}\label{eq18}
    {\rho}=\langle a_l^\dagger a_l\rangle=\frac{1}{N}
   \sum_q\frac{1}{2}\big(\frac{A_q}{\sqrt{A^2_q-B^2_q}}-1\big),
\end{equation}
with $A_q$ and $B_q$ defined by Eqs.~(11). The numerical result of the magnon density for CsNiCl$_3$ is $\rho\approx0.49$ at $D=0$, giving the sublattice magnetization $M\approx0.51$. On the other hand, using slightly different parameter $\xi=1.7\times10^{-2}$ from Ref.~\cite{Welz1993}, we obtain $\rho=0.48$, giving $M=0.52$. Both these results compare favorably with  the experimental value of $M=0.53$ \cite{Welz1993}. As mentioned earlier, our microscopic analysis of the longitudinal modes is based on these mangon density fluctuations and there will be two such modes as discussed in details in the following section.

\section{The longitudinal modes of the quasi-1D hexagonal antiferromagnets}

As mentioned before, the longitudinal excitations in a quantum antiferromagnetic system with a N\'eel-like long-ranged order correspond to the fluctuations in the order parameter. Using the fact that the quantum correction in the order parameter is given by the magnon density $\rho$ as discussed previously in Eq.~(17), the longitudinal modes can be considered as the magnon-density waves. By analogy to Feynman's theory on the low-lying excited states of the helium-4 superfluid \cite{Feynman1954}, the longitudinal excitation states can be constructed by employing the magnon-density operators $S^z$, in contrast the transverse spin-wave excitation states constructed by the spin-flip operators $S^\pm$ \cite{Xian2007}. The energy spectra of these longitudinal collective modes can then be easily derived by a formula first employed by Feynman for the famous phonon-roton spectrum of the helium superfluid involving the structure factor of its ground state.

More specifically, following Feynman as described above, the longitudinal excitation state is approximated by applying the magnon density fluctuation operator $X_q$ to the ground state $|\Psi_g\rangle$ as,
\begin{equation}\label{eq19}
|\Psi_e\rangle =X_q|\Psi_g\rangle,
\end{equation}
where $X_q$ is given by the Fourier transformation of $S^z$ operators,
\begin{equation}\label{eq20}
X_q = \frac{1}{\sqrt{N}}\sum_l e^{i\mathbf {q\cdot r}_l} S^z_l,\quad q>0,
\end{equation}
with index $l$ running over all lattice sites. The condition $q>0$ in Eq.~(20) ensures the orthogonality to the ground state. The energy spectrum for the trial excitation state of Eq.~(19) can be written as
\[
E(q) = \frac{\langle\Psi_g|\tilde X_q HX_q|\Psi_g\rangle}
 {\langle\Psi_e|\Psi_e\rangle}-E_g
 =\frac{\langle\Psi_g|\tilde X_q [H,\,X_q]|\Psi_g\rangle}{\langle\Psi_e|\Psi_e\rangle}
\]
where $\tilde X_q$ is the Hermitian of $X_q$ and where in the second equation we have used the ground state equation, $H|\Psi_g\rangle =E_g|\Psi_g\rangle$. We notice that operator $S^z_l$ in $X_q$ of Eq.~(20) is a Hermitian operator, hence $\tilde X_q=X_{-q}$. By considering the similar excitation state $X_{-q}|\Psi_g\rangle$ with the energy spectrum $E(-q)=E(q)$, it is straightforward to derive \cite{Xian2007},
\begin{equation}\label{eq21}
    E(q)=\frac{N(q)}{S(q)},
\end{equation}
where $N(q)$ is given by the ground-state expectation value of a double commutator as
\begin{equation}\label{eq22}
    N(q)=\frac{1}{2}\langle[X_{-q},[H,X_q]]\rangle_g,
\end{equation}
and the state normalization integral $S(q)$ is the structure factor of the lattice model
\begin{equation}\label{eq23}
    S(q)=\langle\Psi_e|\Psi_e\rangle=\frac{1}{N}\sum_{l,l'}e^{i\mathbf q.(\mathbf r_{l}-\mathbf r_{l'})}\langle S_l^zS_{l'}^z\rangle_g.
\end{equation}
The notation $\langle\dots\rangle_g$ in Eqs.~\eqref{eq22} and \eqref{eq23} indicates the ground-state expectation.

In fact, the excitation state of Eqs.~(19) and (20) can also be viewed as the single-mode approximation (SMA) \cite{Girvin1986} and the expression for $E(q)$ is actually the exact first moment of the dynamic longitudinal structure factor. We also like to point out that the relation between the longitudinal magnon-density waves and the quasiparticle magnon modes can be examined by the first commutation of the operator ~(20) with the Hamiltonian and that the magnon-density waves represent the coherent motion of the spin $\pm1$ magnon pairs, very similar to the plasmon excitations in the electronic systems with coherent motion of quasi-electron-hole pairs as discussed in details in our earlier paper \cite{Xian2006}.

Further support for the form of Eq.~(19) can also be obtained by examining the general structures of the ground and excited states within the framework of the coupled-cluster method (CCM) \cite{Coester1960477,cizek1966,Bishop1991}. Briefly, within the CCM, the ground state is given by applying an exponentiated correlation operator $\hat S$ on a reference state $|\Phi\rangle$ (i.e., the classical N\'eel state in our case) as
\begin{equation}
|\Psi_g\rangle =e^{\hat S}|\Phi\rangle,\quad \hat S=\sum_I F_IC^\dagger_I
\end{equation}
with the multiparticle creation operator $C^\dagger_I$ and the corresponding variational coefficients $F_I$. In our case here, $C^\dagger_I$ is given by the products of the spin-flip operators $S^+$ with respect the N\'eel state. The excitation state within the CCM is given by the linear form as \cite{Emrich1981379,PhysRevB.43.13782,Xian2007}
\begin{equation}
|\Psi_e\rangle =X|\Psi_g\rangle=Xe^{\hat S}|\Phi\rangle,\quad X=\sum_Ix_IC^\dagger_I
\end{equation}
with the variational coefficients $x_I$. In fact, the spin-wave ground state as discussed in Sec.~II can be deduced by a low-order, the so-called SUB2 approximation involving the two-body correlations, in the large-$s$ limit of the CCM \cite{Xian2002}. Furthermore, using the following algebra
\begin{eqnarray*}
S^z_le^{\hat S}&=&e^{\hat S}\bar{S^z_l},\\
\bar{S^z_l}&=&e^{{-\hat S}}S^z_le^{\hat S}=S^z_l+[S^z_l,{\hat S}]
+\frac1{2!}[[S^z_l,{\hat S}],{\hat S}]+\cdots
\end{eqnarray*}
where the nested commutation series in $\bar{S^z_l}$ terminates at the first order in our case, it is not difficult to show the similarity between the excitation state of Eqs.~(19-20) and that of Eq.~(25). The clear advantage of Eqs.~(19) and (20) lies on its simple form and on the fact that the double commutation in $N(q)$ of Eq.~(22) reduces the order of calculations. Furthermore, it satisfies the sum rule as described above in the (SMA).

We have applied these formulas to the bipartite quasi-1D antiferromagnetic systems such as KCuF${}_3$ \cite{yang2011}. For the hexagonal lattice systems as discussed here, we expect that there are two longitudinal modes due to the noncollinear nature of the order parameter on the triangular basal plane. Within the one-boson approach after the two spin rotations as employed here, the two longitudinal modes with $xz$-polarizations of the hexagonal systems can be obtained by folding of the wavevectors in the energy spectra of Eq.~(21), in similar fashion to the one-boson spin-wave theory as discussed in Sec.~II and also to that of Ref.~\cite{Affleck1989}.

Using the Hamiltonian of Eqs.~\eqref{eq5}, it is straightforward to derive the following double commutator with zero anisotropy (i.e. $D=0$) as
\begin{equation}\label{eq25}
\begin{split}
   N(q)=2sJ\sum_\varrho(1+\cos q_z)\tilde g_\varrho
   &+\frac{1}{2}J's\sum_{\varrho'}\Big[
   3(1+\gamma_q)\tilde g_{\varrho'}\\
   &-(1-\gamma_q)\tilde g_{\varrho'}'\Big],
\end{split}
\end{equation}
where $\gamma_q$ is as defined in Eq.~\eqref{eq12} and the transverse correlation functions $\tilde g_\varrho$ and $\tilde g'_\varrho$ are defined respectively as
\begin{equation}\label{eq26}
\tilde g_\varrho=\frac{1}{2s}\langle S_l^+S_{l+\varrho}^+\rangle_g \quad \quad
\tilde g'_\varrho=\frac{1}{2s}\langle S_l^+S_{l+\varrho}^-\rangle_g,
\end{equation}
all independent of index $l$ due to the lattice translational symmetry. Also, the contribution from the three-boson operators with $\sin(\theta_l-\theta_{l+\varrho})$ (the so-called cubic term) is zero. We notice that this cubic term has been included in perturbation theory for the correction in spin-wave spectrum \cite{Miyake1992,Chubukov1994}. In evaluating $\tilde g_\varrho$ and $\tilde g_\varrho'$ of Eq.~\eqref{eq26}, we keep up to the second order in the large-$s$ expansions and obtain
\begin{equation}\label{eq26.1}
\begin{split}
&\tilde g_\varrho=\Delta_\varrho-\frac{2\rho\,\Delta_\varrho+\mu_\varrho\,\delta}{2s},\\
&\tilde g'_\varrho=\mu_{\varrho}-\frac{2\rho \,\mu_{\varrho}+\Delta_{\varrho}\delta}{2s},
\end{split}
\end{equation}
where
\begin{equation}\label{eq27}
\begin{split}
&\rho=\langle a_l^\dagger a_l\rangle=\frac{1}{N}\sum_q\rho_q,\quad\mu_\varrho=\langle a_l^\dagger a_{l+\varrho}\rangle=\frac{1}{N}\sum_qe^{i\bf q\cdot\varrho}\rho_q, \\
& \Delta_\varrho=\langle a_l a_{l+\varrho}\rangle=\frac{1}{N}\sum_qe^{i\bf q\cdot\varrho}\Delta_q,\quad\delta=\langle a_l a_l\rangle=\frac{1}{N}\sum_q\Delta_q,
\end{split}
\end{equation}
and where
\begin{equation}\label{eq27}
\Delta_q=\frac{1}{2}\frac{B_q}{\sqrt{A_q^2-B_q^2}},\quad \rho_q=\frac{1}{2}\big(\frac{A_q}{\sqrt{A_q^2-B_q^2}}
    -1\big),
\end{equation}
with $A_q$ and $B_q$ as given before by Eqs.~\eqref{eq11}. The structure factor within the linear spin-wave approximation is independent of $s$, and is given by
\begin{equation}\label{eq28}
    S(q)=\rho+\frac{1}{N}\sum_{q'}\rho_{q'}\rho_{q+q'}+\frac{1}{N}
    \sum_{q'}\Delta_{q'}\Delta_{q+q'}.
\end{equation}
We like to point out that, the calculations of both Eqs.~\eqref{eq26.1} and \eqref{eq28} involve up to four-boson operators.

We first discuss the general behaviors of the longitudinal spectrum of Eq.~(21) as a function of the ratio of the two nearest-neighbor coupling constants, $\xi=J'/J$. In the limit $\xi\to0$, the Hamiltonian of ~(1) becomes the pure 1d systems; the longitudinal spectrum is gapless and identical to the doublet spin-wave spectra thus forming a triplet excitation state as discussed in details Ref.~\cite{yang2011}. This demonstrates the limitation by the spin-wave ground-state employed, particularly when applied to the integer-spin Heisenberg chain where the Haldane gap is expected as discussed in Sec.~I. In the other limit, $\xi\to\infty$, the Hamiltonian is a pure triangular antiferromagnet with the quasi-gapped longitudinal modes as discussed in details in our previous paper \cite{M.Merdan2012} where we keep only the first order term in Eq.~\eqref{eq26.1} in the large $s$-expansion, similar to the case of the square lattice model.

\begin{figure}
\centering
\includegraphics[scale=0.5]{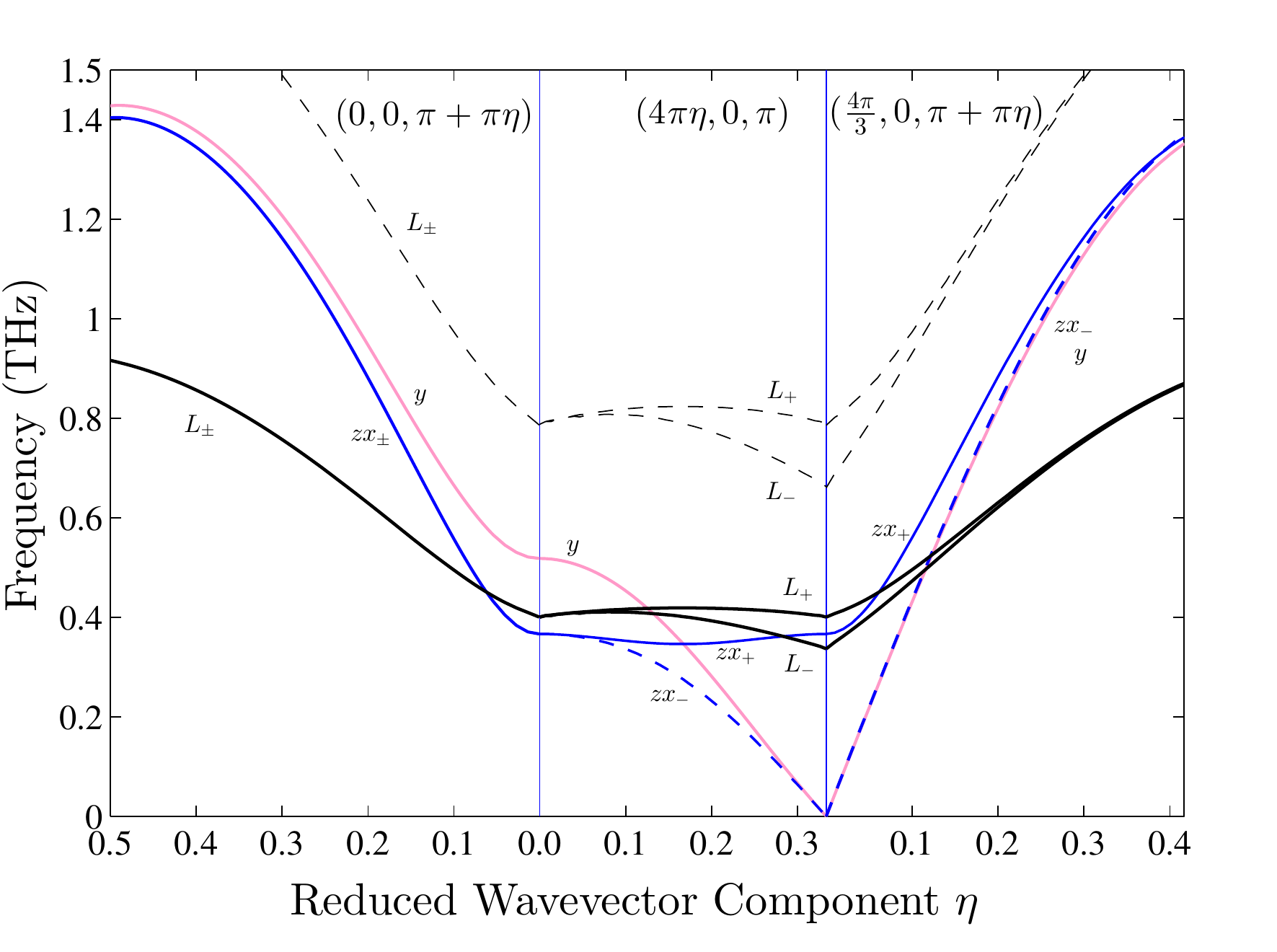}
\caption{The longitudinal modes $L_\pm$ as derived from Eq.~\eqref{eq21} together with the spin-wave $y$- and $zx_\pm$ modes as derived from Eq.~\eqref{eq14} for CsNiCl$_3$ along the symmetry direction $(0,0,\pi+\pi \eta)$, $(4\pi \eta,0,\pi)$ and $(\frac{4\pi}{3},0,\pi+\pi \eta)$. The longitudinal modes $L_\pm$ calculated from the first order approximation and after including the second term in Eq.~\eqref{eq26.1} are indicated by the dash and solid lines respectively.}
\label{fig4}
\end{figure}

\begin{figure}[h]
\centering
\includegraphics[scale=0.5]{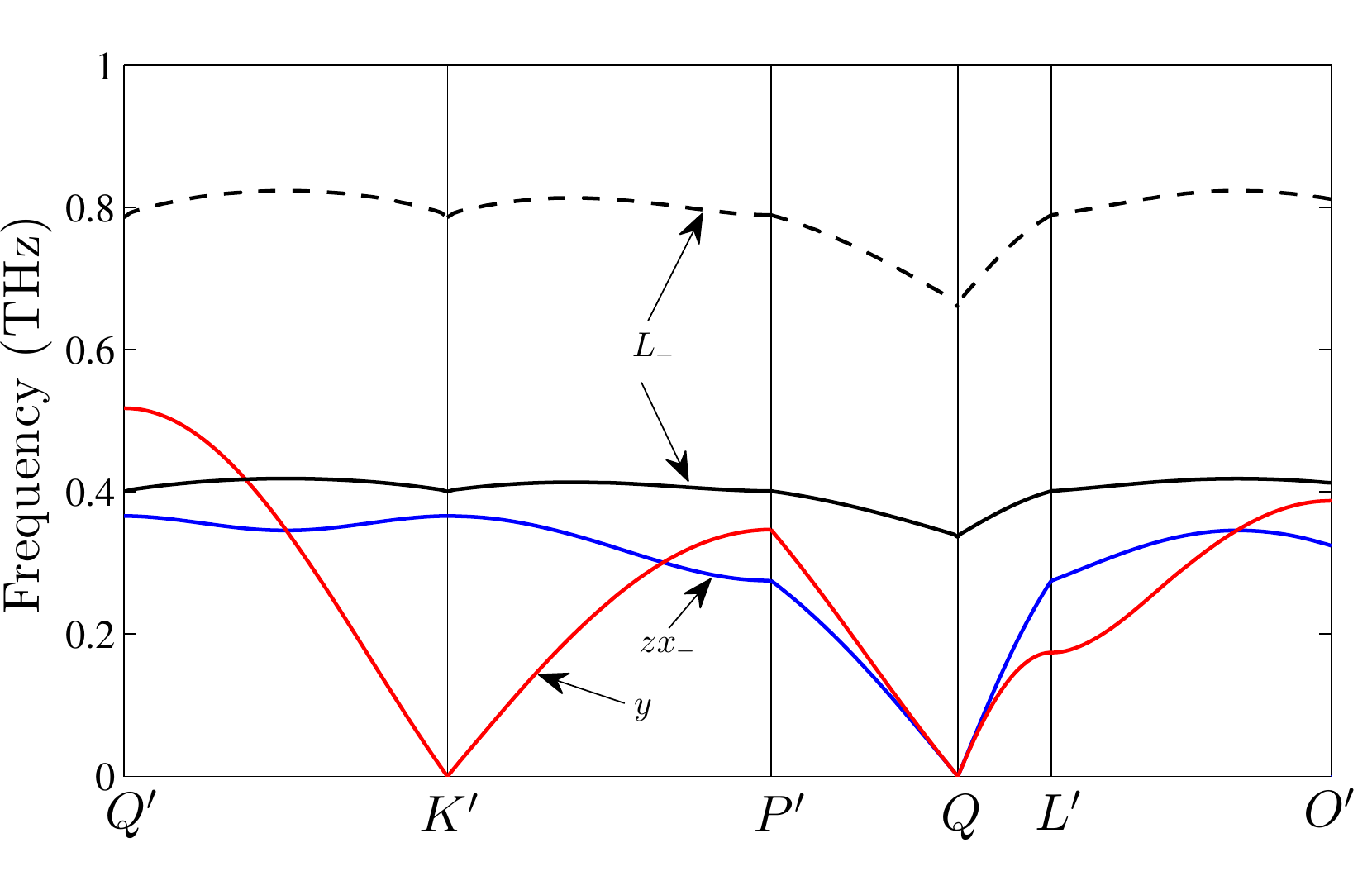}
\caption{The longitudinal mode $L_-$ along path $Q'K'P'QL'O'$ of the hexagonal Brillouin zone of Fig.~\ref{fig2}(b)  together with the spin-wave $y$ and $zx_-$ modes for CsNiCl$_3$. The longitudinal modes $L_\pm$ calculated from the first order approximation and after including the second term in Eq.~\eqref{eq26.1} are indicated by the dash and solid lines respectively.}
\label{fig5}
\end{figure}

For the quasi-1D materials with intermediate values of $\xi$, the spin-wave ground state is a reasonable approximation. We obtain nonzero energy gaps for the longitudinal excitation spectra of Eq.~(21). As discussed before, following Affleck \cite{Affleck1989,PhysRevB.46.8934}, two longitudinal modes for the quasi-1D hexagonal antiferromagnets can be obtained by folding of the wavevector. We denote one as $L_-$ with the spectrum $E(q-Q)$ and the other as $L_+$ with the spectrum $E(q+Q)$. We plot these two longitudinal spectra in the first and second order approximations together with the three spin-wave spectra of Eq.~(15) in Fig.~4 near the magnetic wavector $Q$ for the compound CsNiCl$_3$. Our numerical result for the energy gap of the lower longitudinal mode $L_-$ at $Q$ is $0.96(2J)$ in the first order approximation in Eq.~\eqref{eq26.1}. After including the second order terms the energy gap value is now $(0.49)2J$, in agreement with the experimental results of $0.41(2J)$. We also notice that the upper mode $L_+$ is higher than the $L_-$ mode by about $(0.092)2J$ at $Q$. We also plot the $L_-$ mode along the path $Q'K'P'QL'O'$ of the hexagonal Brillouin zone in Fig.~\ref{fig5} together with the spin-wave $y$ and $zx_-$ modes. As can be seen, the longitudinal mode is nearly flat over the whole spectrum.

For the compound RbNiCl$_3$ also with $s=1$, using the exchange parameters $J=0.485$ and $J'=0.0143$ THz with a larger ratio $\xi=J'/J=0.0295$ \cite{PhysRevB.43.13331}, we obtain similar longitudinal modes as those of CsNiCl$_3$. The numerical result for the energy gap of the $L_-$  mode is 1.16 THz in the first order approximation and 0.69 THz after including the second order contributions at the magnetic wavevector. This later result is in better agreement with the experimental result of about 0.51 THz. We like to point out that there is some difficulty in fitting of Affleck's model with the experimental results for RbNiCl$_3$ \cite{PhysRevB.46.8934,PhysRevB.43.13331}.

Finally we turn to the longitudinal modes for the non-integer-spin quasi-1D hexagonal systems. The superexchange interactions in the hexagonal compound CsMnI${}_3$ can be described by the Hamiltonian of ~(1) with spin quantum number $s=5/2$ and the nearest-neighbor coupling constants $J=0.198$ and $J'=0.001$ THz and negligible anisotropy \cite{PhysRevB.43.679}. This system is very close to the pure 1d system with a very small ratio $\xi=J'/J\approx0.005$. The linear spin-wave theory may be a poor approximation for such a system. Nevertheless, with a similar analysis as before based on the spin-wave ground state, we obtain the $L_-$ mode energy gap value of $0.64$ THz  at the magnetic wavevector $Q$ in the first order approximation, and of $0.47$ THz after including the second order contributions. This later value is still much larger than the experimental value of about $0.1$ THz by Harrison \emph{et al} \cite{PhysRevB.43.679}, which was used to fit a modified spin-wave theory by Plumer and Cail\'e \cite{Plumer1992}. Clearly, for such systems as CsMnI${}_3$, we need a better ground state than that of the spin-wave theory in our analysis.

\section{Discussion}

In this paper, we have investigated the excitation states of the quasi-1D hexagonal systems as modeled by the anisotropic Heisenberg Hamiltonian with only the nearest-neighbor couplings. We have obtained the three spin-wave modes and two longitudinal modes. The energy gaps due to the anisotropy and the energy gaps of the longitudinal modes at the magnetic wavevector are investigated and compared with the experimental results for several quasi-1D hexagonal compounds. We have also estimated the higher-order contributions in the large-$s$ expansions for the longitudinal energy spectra. We like to emphasize that our analysis applies to both integer and non-integer spin systems and there are no other fitting parameters than the nearest-neighbor coupling constants and the anisotropy parameter in the model Hamiltonian provided by experiments. Therefore, the good agreement for the longitudinal energy gap values between our calculations and the experimental measurements for the compounds CsNiCl${}_3$ and RbNiCl${}_3$ are particularly satisfactory. The compound CsMnI${}_3$ is very close to the pure 1d model (i.e., $\xi$ very small) for which the spin-wave ground state is not reliable. It is therefore not surprising to find big discrepancy between our estimate based on spin-wave ground state and the experimental result even after including the higher-order contributions in our calculations. Further improvement may be found on two fronts. Firstly, the contribution of the cubic term may be calculated by a perturbation theory in a similar fashion as employed in Ref.~\cite{Feynman1956,Jackson1962}. Secondly, a better ground state is needed, particularly for the compound CsMnI${}_3$ where the interchain coupling is particularly weak. A more sophisticated many-body theory such as the coupled-cluster method, particularly its recent variational version \cite{Xian2002,Xian2008}, may provide such an improvement.

\bibliographystyle{h-physrev3}

\end{document}